\documentstyle{mn}

\begin{document}

\def\begeq{\begin{equation}}
\def\endeq{\end{equation}}
\def\begeqar{\begin{eqnarray}}
\def\endeqar{\end{eqnarray}}
\def\mf{magnetic field}
\def\MHD{magneto\-hydro\-dyna\-mic}
\def\Amp{Amp\`ere}
\def\elm{electro\-magne\-tic}
\def\magn{mag\-ne\-tic}
\def\dy{dynamo}
\def\gm{gra\-vi\-to\-\magn}
\def\po{po\-lo\-idal}
\def\tor{to\-ro\-idal}
\def\Gauss{\rm Gauss}
\def\s{\rm s}
\def\yr{\rm yr}
\def\K{\rm K}
\def\cm{\rm cm}
\def\eV{\rm eV}
\def\Msol{M_\odot}
\def\BL{Boyer\--Lindquist}
\def\equ{equation}
\def\1d#1{{1\over#1}}
\def\del{\partial}
\def\delt#1{{\del#1 \over \del t}}
\def\delthe#1{{\del#1 \over \del\theta}}
\def\delr#1{{\del#1 \over \del r}}
\def\ddtau#1#2{{d#2 \over d\tau_{\rm #1}}}
\def\ti{\tilde}
\def\pri{^{\, '}}
\def\rmi{^{\rm i}}
\def\rme{^{\rm e}}
\def\rmx{^{\rm x}}
\def\o{\omega}
\def\ot{\tilde \omega}
\def\a{\alpha}
\def\ag{\alpha_{\rm g}}
\def\rg{r_{\rm g}}
\def\vbeta{\vec{\beta}}
\def\phid{\hat\phi}
\def\E{\vec{E}}
\def\Ep{\vec{E}_{\rm p}}
\def\Eph{E^{\phi}}
\def\Ephid{E^{\phid}}
\def\B{\vec{B}}
\def\Bp{\vec{B}_{\rm p}}
\def\Bph{B^{\phi}}
\def\Bphid{B^{\phid}}
\def\jph{j^{\phi}}
\def\jphid{j^{\phid}}
\def\j{\vec{j}}
\def\jp{\vec{j}_{\rm p}}
\def\v{\vec{v}}
\def\vp{\vec{v}_{\rm p}}
\def\S{\vec{S}}
\def\et{\vec{e}_{t}}
\def\ephi{\vec{e}_{\phi}}
\def\ephid{\vec{e}_{\hat\phi}}
\def\nonu{\nonumber}
\def\n{\vec{\nabla}}
\def\rot{\n \times}
\def\div{\n\cdot}
\def\tarrow#1{\buildrel\leftrightarrow\over #1}
\def\tens#1{\ifmmode\mathchoice{\mbox{$\sf\displaystyle#1$}}
{\mbox{$\sf\textstyle#1$}}
{\mbox{$\sf\scriptstyle#1$}}
{\mbox{$\sf\scriptscriptstyle#1$}}\else
\hbox{$\sf\textstyle#1$}\fi}
\newcommand{\subi}{_{\rm i}}
\newcommand{\sube}{_{\rm e}}
\newcommand{\subx}{_{\rm x}}
\newcommand{\mi}{m_{\rm i}}
\newcommand{\me}{m_{\rm e}}
\newcommand{\mx}{m_{\rm x}}
\newcommand{\rhomi}{\rho_{\rm mi}}
\newcommand{\rhome}{\rho_{\rm me}}
\newcommand{\rhomx}{\rho_{\rm mx}}
\newcommand{\rhom}{\rho_{\rm m}}
\newcommand{\rhoci}{\rho_{\rm ci}}
\newcommand{\rhoce}{\rho_{\rm ce}}
\newcommand{\rhoc}{\rho_{\rm c}}
\newcommand{\gammai}{\gamma_{\rm i}}
\newcommand{\gammae}{\gamma_{\rm e}}
\newcommand{\gammax}{\gamma_{\rm x}}

\title[MHD description of a two-component plasma in the Kerr metric]{On the 
magneto\-hydro\-dynamic description of a two-component plasma in 
the Kerr metric} 
\author[Ramon Khanna]
{Ramon Khanna\\
Landessternwarte, K\"onigstuhl, D-69117 Heidelberg, Germany\\
rkhanna@lsw.uni-heidelberg.de}

\maketitle

\begin{abstract}
The magnetohydrodynamic \equ s describing an inviscid fully ionized plas\-ma in 
the vicinity of a rotating black hole are derived from a two-component plasma 
theory within the framework of the 3+1 split of the Kerr metric.
Of central interest is the generalized Ohm's law. In the limit of quasi-neutral 
plasma it contains no new terms as compared with special relativity. 
Gravitomagnetic terms appear in Ohm's law only, if the plasma is charged in its 
rest frame or the MHD-approximation is not applied.

It is argued that a relativistic single-fluid description of a 
multiple-com\-po\-nent plasma is possible only for cold 
(i.e. intrinsically non-relativistic) components. As seen by local stationary 
observers, close to the horizon the electron collision time becomes longer than 
dynamical timescales, i.e. the plasma appears to behave as particles.
\end{abstract}


\section{INTRODUCTION}
Astrophysics of black hole systems with their magnetized accretion discs, 
magnetospheres and jets has been a fascinating field of research for some 
twenty years. There are a number of papers dealing with the 
general-relativistic description of those systems. The majority of the papers 
is restricted to ideal MHD of black hole magnetospheres and jets 
(Blandford \& Znajek 1977; Phinney 1983; Bekenstein \& Eichler 
1985; Camenzind 1986, 1987; Takahashi et al. 1990; Okamoto 1992; 
Beskin \& Par'ev 1993; Fendt 1997). 
General-relativistic resistive MHD has been discussed by Bekenstein \& Oron 
(1978). Khanna \& Camenzind (1994, 1996a,b), Khanna (1997) and 
Kudoh \& Kaburaki (1996) have 
applied resistive MHD to accretion discs of Kerr black holes. In both 
ideal and resistive MHD the authors have used the standard Ohm's law.

Generalized Ohm's laws in special-relativistic context have been derived 
from invariant distribution functions by Ardavan (1976) and by 
Blackman \& Field (1993). Ardavan shows that the simple ideal MHD condition 
from the standard Ohm's law breaks down in extreme regions of a pulsar 
magnetosphere.
This result suggests that also around a black hole there should be regions 
in which the standard Ohm's law is merely a poor approximation.

In this paper I derive the full set of MHD equations, including a generalized 
Ohm's law, for an inviscid, fully ionized plasma from a two-component plasma 
theory within the framework of the 3+1 split of the Kerr metric. I discuss 
two major issues: (i) under which circumstances is a relativistic single-fluid 
description of a multiple-component plasma possible? 
(ii) Does the \gm\ field play a role in Ohm's law? 

The \equ s and results of Sect.~2 are general and apply also to an 
electron-positron plasma. The MHD description of Sect.~3 is limited to 
a relativistically cold electron-ion plasma.

I use a metric with signs \((-{}+{}+{}+)\) and set $c=1$. 
Spacetime vectors and tensors are written upright. In 3-dimensional space 
vectors and tensors are denoted by arrows (e.g. $\v$ and $\tarrow{T}$) and 
covariant derivatives are indicated by vertical bars, 
e.g. \(\nabla_j \beta^i \equiv \beta^i_{\; |j}\). Superscripts i, e, $\pri$ 
indicate the rest frames $K\rmi$, $K\rme$ and $K\pri$, in which a quantity is 
defined.

\section{Two-component plasma theory in the 3+1 split of the Kerr metric}
In this section I will first discuss the relativistic definition of a 
plasma as the centre-of-mass fluid of its two components and introduce the 
various rest frames involved. Then I will summarize the 3+1 split of the Kerr 
metric, in which the equations governing 
the individual components of the plasma will be written down and 
finally combined (in Sect.~\ref{MHD}) to obtain the 
equation of motion, the local law of energy conservation and the generalized 
Ohm's law for the plasma as single fluid.

\subsection{Basic concepts, definitions and \equ s}
\subsubsection{Ions, electrons and plasma}
\label{secIep}
A fluid of charged particles is described by its stress-energy 
spacetime tensor, 
its number-flux 4-vector and its current 4-vector $\tens{J}_{\rm x}$. 
Thus for interacting ions (in this section `ions' could mean positrons 
as well) and electrons 
we have for the hydrodynamic part of the stress-energy spacetime-tensor
\begeq
        T^{\a\beta}_{\rm x} = 
        (\rho_{\rm mx}^{\rm x} + p_{\rm x}^{\rm x})
	W_{\rm x}^{\a}W_{\rm x}^{\beta} + p_{\rm x}^{\rm x} g^{\a\beta}
	+ T^{\a\beta}_{\rm x\, coll}\; ,
\label{T4x}
\endeq
where \(\rho_{\rm mx}^{\rm x} = n_{\rm x} m_{\rm x} (1 + \Pi)\) is the total 
density of mass-energy (including specific internal energy $\Pi$) and
$p_{\rm x}^{\rm x}$ is the total pressure of the species `x' as measured in 
the rest frame $K^{\rm x}$ of the species (x=i,e). $\tens{W}_{\rm x}$ is 
the bulk 4-velocity of the species. The interaction between electrons and 
ions is 
described by $T^{\a\beta}_{\rm x\, coll}$. The number-flux 4-vectors are 
\begeq
	N\subx^{\a} = n_{\rm x} W_{\rm x}^{\a}\; ,
\endeq
with the species' rest frame particle densities $n_{\rm x}$.

The plasma is assumed to be a perfect fluid and is defined by the sums of the 
ion and electron tensors:
\begeq
        (\rho_{\rm m}\pri + p\pri)W^{\a}W^{\beta} + p\pri g^{\a\beta}
        \equiv T^{\a\beta} = T^{\a\beta}\subi + T^{\a\beta}\sube \; ,
\label{defplas}
\endeq
\begeq
	N^{\a} = N\subi^{\a} + N\sube^{\a}
\endeq
and 
\begeq
	J^{\a} = J^{\a}\subi + J^{\a}\sube\; .
\endeq
Note that, from Eq.~(\ref{defplas}), the plasma's 4-velocity, energy density, 
momentum density and pressure are 
uniquely defined as centre-of-mass quantities of ions plus electrons. This 
means, in particular, that \( N^{\a} \neq n_{\rm p} W^{\a} \), where 
$n_{\rm p}$ would be the number density of `plasma particles', 
since there is 
no more freedom for such a definition. The consequences of this fact for the 
possibility of describing the plasma as a single fluid will be discussed below. 

Electrodynamic parts of $\tens{T}_{\rm x}$ and $\tens{T}$ are not given  
here, but will appear as Lorentz force and electric energy in the laws of 
momentum conservation and energy conservation below.

\subsubsection{Rest frames}
\label{rframes}
The plasma rest frame \(K\pri\) is defined as the frame in which the 
plasma's centre-of-mass is at rest
\begeq
	\v\pri = \frac{(\rhomi\rmi+p\subi\rmi)\v\subi\pri + 
	(\rhome\rme+p\sube\rme)\v\sube\pri 
	+ \S_{\rm i\, coll}\pri + \S_{\rm e\, coll}\pri}
	{ \rhom\pri+p\pri } = \vec{0}
\label{vpri}
\endeq
(for the definition of momentum densities see below).
The idea of a plasma implies that, in the plasma rest frame, both the 
ion-component (subscript `i') and the electron-component (subscript `e') 
have non-relativistic {\it bulk} velocities 
\(\v_{\rm i}\pri\) and \(\v_{\rm e}\pri\), i.e. the bulk Lorentz factors 
\(\gamma_{\rm i}\pri=1\) and \(\gamma_{\rm e}\pri=1\) (this does not mean that 
the plasma components may not be relativistically hot). In this sense the 
rest frames of electrons $K\rme$, ions $K\rmi$ and plasma $K\pri$ are identical 
and quantities defined in $K\rme$ and $K\rmi$ will be the same in $K\pri$ to 
high accuracy.

The total densities of mass-energy as measured in $K\pri$ are 
\begeqar
	\epsilon\subi\pri&=&(\rhomi\rmi+p\subi\rmi\, v\subi^{\, '\, 2})
		\gamma\subi^{\, '\, 2} + \epsilon_{\rm i\, coll}\pri\approx
	\rhomi\rmi
\nonu\\
        \epsilon\sube\pri&=&(\rhome\rme+p\sube\rme\, v\sube^{\, '\, 2})
                \gamma\sube^{\, '\, 2} + \epsilon_{\rm e\, coll}\pri\approx
        \rhome\rme
\nonu\\
        \epsilon\pri&=&\rhom\pri = \epsilon\subi\pri + \epsilon\sube\pri
        \approx \rhomi\rmi + \rhome\rme \; ,
\label{epspri}
\endeqar
where the approximations are due to \(v\subx^{\, '\, 2}/c^2 \ll 1\) and 
the energy exchange in collisions being small compared to the rest mass of the 
particles. The latter is true for a plasma in equilibrium, i.e. behaving as 
one fluid. In this case collisions between electrons and ions are nearly 
elastic and 
also the energy flux $\S_{\rm x\, coll}\pri$ is negligible.
The pressures \(p_{\rm i}\pri\simeq p_{\rm i}\rmi\) and
\(p_{\rm e}\pri\simeq p_{\rm e}\rme\) add up to the total pressure
\begeq
        p\pri= p_{\rm i}\pri+p_{\rm e}\pri \; .
\endeq
The charge densities are 
\begeq
        \rho_{\rm ci}\pri\simeq \rho_{\rm ci}\rmi= Z e n_{\rm i} \quad 
        \rho_{\rm ce}\pri\simeq \rho_{\rm ce}\rme= - e n_{\rm e} \quad
        \rho_{\rm c}\pri= \rho_{\rm ci}\pri + \rho_{\rm ce}\pri \; .
\label{rhocpri}
\endeq

\subsubsection{The 3+1 split of the Kerr metric}
In the 3+1 split of the Kerr metric (see Thorne et al. 1986), 
spacetime is split into a family of three dimensional differentially 
rotating hypersurfaces of constant time with internal curvature. 
Those hypersurfaces can be mentally collapsed 
into a single 3-dimensional `absolute space' in which time is globally
described by the \BL\ coordinate $t$.
The line element of the Kerr metric in 3+1 notation is given by
\begeqar
        ds^2 &=& g_{\a\beta}dx^{\a}dx^{\beta} \nonu\\
	&=& 
	-\ag^2\, dt^2 + h_{jk}(dx^j + \beta^j\, dt)
                                         (dx^k + \beta^k\, dt)  \; ,
\label{dsKerr31}
\endeqar
where the lapse function is identified with the gravitational redshift
\begeq
        \ag\equiv \left(\frac{d\tau}{dt}\right)_{\rm FIDO} = {\rho\over\Sigma}\sqrt{\Delta}\; ,
\endeq
$\tau$ being the local proper time as measured by FIDOs (see below).
The shift functions are the components of the gravitomagnetic potential 
$\vbeta$
\begeq
        \beta^r = \beta^{\theta} = 0\; ,\quad
            \beta^{\phi}\equiv -\o = -\left(\frac{d\phi}{dt}\right)_{\rm FIDO} = 
	- {2aMr\over\Sigma^2}\; ,
\label{shift}
\endeq
and the components of the 3-metric $\tarrow{h}$ are 
\begeq
        h_{rr} = {\rho^2\over \Delta}\; ,\ \ 
            h_{\theta\theta} = \rho^2\; ,\ \ 
            h_{\phi\phi} = \ot^2\; ,\ \ 
            h_{jk} = 0\; ;(j\ne k)                     \; .
\label{hikKerr}
\endeq
Note that $h_{jk} = g_{jk}$ but $h^{jk}= g^{jk}+\beta^j\beta^k / \ag^2$.
The metric functions appearing here are defined as
\begeqar
        \Delta &\equiv& r^2-2Mr + a^2\; ,\quad
            \rho^2 \equiv r^2+a^2\cos^2{\theta}\; , \nonu\\
        \Sigma^2 &\equiv& (r^2+a^2)^2-a^2\Delta\sin^2\theta\; ,\quad
            \ot \equiv (\Sigma /\rho )\sin\theta \; .
\endeqar
The parameters of the black hole are its mass $M$ and angular momentum 
$J$, which define the Kerr parameter \(a\equiv J/M\).

\subsubsection{Quantities and equations in the FIDO frame}
Fiducial observers (FIDOs) who live in `absolute Kerr-space'
are stationary and locally non-rotating zero angular momentum observers 
(ZAMOs). As seen in a global \BL\ coordinate system, ZAMOs are in shearing 
circular motion around the Kerr black hole, described by their 4-velocity 
\begeq
	\tens{U} = \ag^{-1} (\et - \vbeta)\; .
\endeq

The quantities and equations describing physics in the locally flat 
laboratories of the FIDOs are obtained from spacetime vectors, tensors and 
tensorial equations by projection along \(\tens{U}\) and orthogonal to 
\(\tens{U}\) with the projection operator
\begeq
	h_{\a\beta} = g_{\a\beta} + U_{\a} U_{\beta}\; .
\endeq
The relevant plasma quantities in the context of this paper are, 
for ions and electrons respectively, the bulk 
velocities $\v_{\rm i}$, $\v_{\rm e}$ 
and Lorentz factors $\gamma_{\rm i}$, $\gamma_{\rm e}$, 
the energy densities 
\begeq
        \epsilon\subx=(\rhomx\rmx + p\subx\rmx\v\subx^2)
	\gamma\subx^2 + \epsilon_{\rm x\, coll}
	\approx (\rhomx\rmx + p\subx\rmx\v\subx^2)
        \gamma\subx^2
\label{epsx}
\endeq
and the momentum densities 
\begeq
	\vec{S}\subx=(\rhomx\rmx + p\subx\rmx)\gamma\subx^2
				\v\subx + \S_{\rm x\, coll}
	\approx (\rhomx\rmx + p\subx\rmx)\gamma\subx^2
                                \v\subx	\; .
\label{Sx}
\endeq
The stress-energy tensor for an interacting fluid in `absolute space' 
is given by 
\begeq
	\tarrow{T}_{\rm x} = (\rho_{\rm mx}\rmx + 
	p_{\rm x}\rmx)\gamma_{\rm x}^2 \v_{\rm x}\otimes\v_{\rm x} 
	+ p_{\rm x}\rmx\tarrow{h} + \tarrow{T}_{\rm x\, coll}\; .
\label{Tx}
\endeq
The plasma can be treated as a single fluid, if collisions between ions and 
electrons involve energies and energy fluxes, much smaller than the rest mass 
of the particles. Then the collision 
terms can be neglected in Eqs.~(\ref{epsx}) and (\ref{Sx}). 
In $\tarrow{T}_{\rm x}$ the collision term has to be kept, however, as it 
describes the momentum flux, which couples ions and electrons.
Note that $\epsilon_{\rm x}$, $\vec{S}_{\rm x}$ and 
$\tarrow{T}_{\rm x}$ result from splitting $\tens{T}_{\rm x}$.

The charge densities and current densities of the plasma components are, 
respectively,
\begeq
        \rho_{\rm ci} = \rho_{\rm ci}\rmi\gamma_{\rm i} = 
		\frac{\rho_{\rm ci}\pri} {\gamma (1-\v\cdot\v_{\rm i})} \quad
        \rho_{\rm ce} = \rho_{\rm ce}\rme\gamma_{\rm e} =
		\frac{\rho_{\rm ce}\pri} {\gamma (1-\v\cdot\v_{\rm e})}
\label{rhocie}
\endeq
\begeq
	\j_{\rm i} = \rho_{\rm ci} \v_{\rm i} \qquad
        \j_{\rm e} = \rho_{\rm ce} \v_{\rm e}\; . 
\label{jie}
\endeq
Equations~(\ref{rhocie}) are derived in appendix~\ref{apptransrho}. 

In the context of a two-component theory, each plasma component 
(subscript x=i,e) has to obey the continuity \equ\ for the density of 
rest-mass (see appendix~\ref{appconspart})
\begeq
	\left(\delt{} - \vec{\beta}\cdot\n\right)
	(n\subx m\subx\gamma\subx ) 
	+ \div(\ag n\subx m\subx\gamma\subx \v_{\rm x}) = 0
\label{continuity}
\endeq
and the law of momentum conservation (Thorne et al. 1986; Thorne \& Macdonald 
1982)
\begeqar
	\lefteqn{
	\1d{\ag}\left(\delt{} - \vbeta\cdot\n\right)\vec{S}_{\rm x\, id}=
     \epsilon_{\rm x\, id}\, \vec{g} + \tarrow{H}\cdot\vec{S}_{\rm x\, id}
	}\nonu\\
	&& -{} \1d{\ag}\div(\ag\tarrow{T}_{\rm x\, id})
	+ \rho_{\rm cx}\E + \j_{\rm x}\times\B
	+ \vec{k}_{\rm x} \; ,
\label{momentum}
\endeqar
where 
\begeqar
	\vec{k}_{\rm x}&=&\epsilon_{\rm x\, coll}\, \vec{g} 
	+ \tarrow{H}\cdot\vec{S}_{\rm x\, coll}
        - \1d{\ag}\div(\ag\tarrow{T}_{\rm x\, coll})	\nonu\\
	&-&{}\1d{\ag}\left(\delt{} - \vbeta\cdot\n\right)\vec{S}_{\rm x\, coll}
	\approx - \1d{\ag}\div(\ag\tarrow{T}_{\rm x\, coll})
\endeqar
is an interaction force density due to collisions between ions and electrons. 
Note that, on physical grounds, 
\(	\vec{k}_{\rm i} = - \vec{k}_{\rm e} \; .\) The subscript `id' refers 
to the ideal (inviscid) part of the quantity.
Equation~(\ref{momentum}) contains all the terms known from flat space plus 
an accelerating force along $\vbeta$ on the l.h.s. plus, on the r.h.s., a 
coupling term between the gravitomagnetic tensor field 
\begeq
	\tarrow{H}\equiv \ag^{-1} \n\vbeta\; ,\qquad 
	H_{ij}=\ag^{-1}\beta_{j|i} 
\endeq
and $\S_{\rm x}$, which acts as gravitomagnetic 
force similar to the electromagnetic Lorentz force. The gravitational 
acceleration is related to the gravitational redshift by \(\vec{g}=-\n\ln\ag\).

The \MHD\ law of energy conservation for each individual fluid is 
(Thorne et al. 1986; Thorne \& Macdonald 1982) 
\begeq
        \1d{\ag}\left(\delt{} - \vbeta\cdot\n\right)\epsilon_{\rm x}
        = -\div\S_{\rm x} + 2 \vec{g}\cdot\S_{\rm x} 
        + \tarrow{H}:\tarrow{T}_{\rm x} + \j_{\rm x}\cdot\E\; .
\endeq
The term $\tarrow{H}:\tarrow{T}_{\rm x}\equiv H_{ij}T_{\rm x}^{ij}$ can be 
interpreted as the rate per unit volume at which the \gm\ field does work 
on matter.

Furthermore the electrodynamics of the plasma is 
described by Maxwell's \equ s and the law of charge conservation (Thorne et al. 
1986)
\begeq
	\left(\delt{} - \vec{\beta}\cdot\n\right)\rho_{\rm cx}
        + \div(\ag \j_{\rm x}) = 0\; .
\label{conscharge}
\endeq	

\section{One-fluid description of a cold electron-ion plas\-ma}
\label{MHD}
In this section the two sets of \equ s for ions (not positrons) and electrons 
will be 
combined to yield the \equ s of MHD, which should no longer contain quantities 
of the individual components. It is argued that this is possible only 
for a cold plasma. `Cold' means intrinsically non-relativistic, i.e.
\(p\subx\rmx \ll  \rho_{\rm mx}\rmx c^2\), 
\(\ \rho_{\rm mx}\rmx \approx n\subx m\subx\) (specific internal energy density
negligible compared to rest mass density) and 
\(\rhom\pri\approx n\subi\mi + n\sube\me\). This limit has to 
be taken of Eqs.~(\ref{T4x}) and (\ref{defplas}), or, 
equivalently, in Eqs.~(\ref{vpri}), (\ref{epspri}), (\ref{epsx}) -- (\ref{Tx}).

\subsection{Particle conservation}
Describing a plasma as a single fluid means that one has to define 
`plasma particles' (subscript `p'), which have to fulfil 
particle conservation 
\begeq
        \left(\delt{} - \vbeta\cdot\n\right)(n_{\rm p}\gamma_{\rm p})
        + \div(\ag n_{\rm p}\gamma_{\rm p}\v_{\rm p}) = 0\; ,
\endeq
or 
\(N^{\alpha}_{\ ;\alpha} \equiv (n_{\rm p} W_{\rm p}^{\alpha})_{;\alpha} = 0\).
As was mentioned in Sect.~\ref{secIep}, identifying $\v_{\rm p}$ with the 
centre-of-mass velocity $\v$ (or $W_{\rm p}^{\alpha}$ with $W^{\alpha}$) 
would be in conflict with Def.~(\ref{defplas}).
This means that, without any further assumptions, `plasma particle' 
conservation can not be formulated in terms of $\v$. 
Adding the continuity equations~(\ref{continuity}) (x=i,e) of the two 
components rather than the \equ s of particle conservation yields
\begeqar
	\lefteqn{
        \left(\delt{} - \vbeta\cdot\n\right)(n\subi\mi \gammai 
	+ n\sube\me \gammae )	}\nonu\\
        && +{\, } \div(\ag n\subi\mi\gammai\v\subi + 
	\ag n\sube\me\gammae\v\sube) = 0\; .
\label{plascontie}
\endeqar
We want to eliminate the $\gammai$, $\gammae$, $\v\subi$ and $\v\sube$ in 
accordance with Definition~(\ref{defplas}) and derive a set of \equ s that 
contains only quantities of a one-fluid description.
\(T^{\a\beta} = T^{\a\beta}_{\rm i} + T^{\a\beta}_{\rm e}\) splits into
\begeq
	\epsilon = \epsilon_{\rm i} + \epsilon_{\rm e}
\label{endens}
\endeq
and
\begeq
        \S = \S_{\rm i} + \S_{\rm e}\; .
\label{momdens}
\endeq
Definition~(\ref{defplas}) implies that 
\begeq
        \epsilon \equiv (\rho_{\rm m}\pri + p\pri v^2)\gamma^2 
	\approx \rho_{\rm m}\pri \gamma^2\; ,
\label{epsdef}
\endeq
\begeq
        \S \equiv (\rho_{\rm m}\pri + p\pri)\gamma^2\v\approx \rho_{\rm m}\pri 
	\gamma^2\v 
\label{Sdef}
\endeq
like obviously a FIDO would define energy density and momentum density of a 
plasma with velocity $\v$ and Lorentz factor $\gamma$ anyway. 
In the cold plasma limit and with \(\gammae\approx\gammai\approx\gamma\),
the last five \equ s, together with Eqs.~(\ref{epsx}) and (\ref{Sx}), yield
\begeq
        \left(\delt{} - \vbeta\cdot\n\right)(\rhom\pri\gamma)
        + \div(\ag \rhom\pri\gamma\v)= 0\; .
\label{contplas}
\endeq
It is important to note that this continuity \equ\ for the plasma, 
expressed in terms of centre-of-mass quantities, could only be derived 
in the cold plasma limit. This suggests (though a strict proof has not been 
given) that a hot multiple-component plasma 
may not be described as a single fluid at all. This has far reaching 
consequences for modelling hot astrophysical plasmas in jets and in hot 
transsonic accretion flows onto black holes.

Note that \(\gammae\approx\gammai\) is not really a new assumption, but 
is a consequence of \(\gammae\pri=\gammai\pri = 1\). Furthermore, with the 
cold plasma assumption and \(n\subi\mi \gg n\sube\me\), 
Eqs.~(\ref{endens}), (\ref{epsdef}) and Eqs.~(\ref{momdens}), (\ref{Sdef}) 
imply
\begeq
        \gamma^2\approx\gamma_{\rm i}^2 \hbox{ and }
        \gamma^2\v\approx \gamma_{\rm i}^2\v_{\rm i}\; .
\label{gamma2v}
\endeq

\subsection{Equation of motion}
\label{seceqmot}
In the cold plasma limit $\tarrow{T}_{\rm x\, id}$ 
can be eliminated from Eq.~(\ref{momentum}), using Eq.~(\ref{continuity}),
to yield the \equ\ of motion for each plasma component:
\begeqar
	\lefteqn{
	\rhomx\rmx\gamma\subx\left[\1d{\ag}\delt{}+\left(\v\subx
	-\frac{\vbeta}{\ag}\right)\cdot\n\right](\gamma_{\rm x}\v\subx)
	 = \rho_{\rm mx}\rmx\gamma_{\rm x}^2\vec{g} 	}\nonu\\
	&& +{} \rho_{\rm mx}\rmx\gamma_{\rm x}^2\tarrow{H}\cdot\v_{\rm x} -
	\1d{\ag}\n(\ag p_{\rm x}\rmx)
	+ \rho_{\rm cx}\E + \j_{\rm x}\times\B + 
	\vec{k}_{\rm x}\; .
\label{eqmotx}
\endeqar
The \equ\ of motion for an inviscid electron-ion plasma in one-fluid 
description follows by adding Eqs.~(\ref{eqmotx}) (x=i,e):
\begeqar
	\lefteqn{
        \rhom\pri\gamma\left[\1d{\ag}\delt{}+\left(\v
        -\frac{\vbeta}{\ag}\right)\cdot\n\right](\gamma\v)
        = \rhom\pri\gamma^2\vec{g} }\nonu\\
	&&+{} \rhom\pri\gamma^2\tarrow{H}\cdot\v -
        \1d{\ag}\n(\ag p\pri)
        + \rho_{\rm c}\E + \j\times\B \; .
\label{eqmot1}
\endeqar
It does not come as a surprise that this \equ\ looks just like 
Eq.~(\ref{eqmotx}). This is, however, not trivial and includes definitions and 
approximations that require thorough discussion, in particular in the 
relativistic context. This will be the topic of the rest of this section.
Note that Eq.~(\ref{eqmot1}) could be derived straightforwardly by adding up 
the momentum \equ s (see Sect.~\ref{secmom} below) and eliminating 
\(\tarrow{T} = \tarrow{T}\subi + \tarrow{T}\sube\) 
with the aid of Eq.~(\ref{contplas}). This way, however, the physics behind the 
assumptions that have to be made (or that have been made already) 
would remain obscure.

As has been elaborated in the last section, energy densities and momentum 
densities add according to Eq.~(\ref{defplas}).
Adding the current density 4-vectors of each plasma component and 
splitting into charge density and current density yields 
\begeq
	\rho_{\rm c} \equiv \rho_{\rm ci} + \rho_{\rm ce} = 
	Z e n_{\rm i}\gamma_{\rm i} - e n_{\rm e}\gamma_{\rm e}
	\; ,
\endeq
\begeq
	\j \equiv \j_{\rm i} + \j_{\rm e} = 
	Z e n_{\rm i}\gamma_{\rm i}\v_{\rm i} 
	- e n_{\rm e}\gamma_{\rm e}\v_{\rm e}
    \; .
\endeq
The expressions for $ \rho_{\rm c}$ and $\j$ involve Eqs.~(\ref{rhocpri}), 
(\ref{rhocie}) and (\ref{jie}). Note that the collision force densities have 
cancelled in Eq.~(\ref{eqmot1}).

Besides using the definitions above, there is a fundamental 
assumption that has been made to arrive at the l.h.s. of Eq.~(\ref{eqmot1}). 
It has been assumed that the two plasma components are coupled sufficiently 
strong that their bulk accelerations 
\begeq
	\ddtau{x}{(\gamma_{\rm x}\v_{\rm x})}\equiv
	\left[\frac{\gamma_{\rm x}}{\ag}\delt{}
	+\gamma_{\rm x}\left(\v_{\rm x}
        -\frac{\vbeta}{\ag}\right)\cdot\n\right](\gamma_{\rm x}\v_{\rm x})
\label{bulkacc}
\endeq
are nearly the same (standard assumption for the applicability of MHD), i.e. 
\begeq
	\left|\ddtau{i}{(\gamma_{\rm i}\v_{\rm i})} 
	- \ddtau{e}{(\gamma_{\rm e}\v_{\rm e})}\right|
	 \ll 
	\left|\ddtau{i}{(\gamma_{\rm i}\v_{\rm i})}\right|\; ,
\label{schldiff}
\endeq
where $\tau_{\rm x}$ is the proper time in the frame $K\rmx$.
The application of this assumption (see Schl\"uter 1950; Kippenhahn \& 
M\"ollenhoff 1975 for the non-relativistic equivalent) in the derivation of
Eq.~(\ref{eqmot1}) is elaborated in appendix~(\ref{appschldiff}).
In Kerr metric we have to go even a step further for consistency. The 
acceleration due to the \gm\ 
force is of the same order as the acceleration along $\vbeta$. In fact, if 
both \gm\ terms of Eq.~(\ref{eqmotx}) are added together their axisymmetric 
\tor\ components cancel. If inequality~(\ref{schldiff}) were
applied alone in the derivation of the generalized Ohm's law (see below and 
appendix~\ref{fullohm}), such an axisymmetric \tor\ term  would remain as 
an artefact \dy\ term.
We therefore also have to assume that the coupling between ions and electrons 
synchronizes the \gm\ accelerations, i.e.
\begeq
	\left|\tarrow{H}\cdot(\gammai^2\v\subi) - 
	\tarrow{H}\cdot(\gammae^2\v\sube)\right|
         \ll
        \left|\tarrow{H}\cdot(\gammai^2\v\subi)\right|\; .
\label{syncHv}
\endeq
This assumption has no effect on the \equ\ of motion, but is crucial in the 
derivation of Ohm's law. 

\subsubsection{Exact law of momentum conservation for a two-component 
plasma}\label{secmom}
As was mentioned above, Eqs.~(\ref{momentum}) (with x=i,e) can be added to 
give the exact law of momentum conservation for a two-component plasma
\begeqar
        \1d{\ag}\left(\delt{} - \vbeta\cdot\n\right)\vec{S}&=&
     \epsilon\vec{g} + \tarrow{H}\cdot\vec{S}
        - \1d{\ag}\div(\ag\tarrow{T})	\nonu\\
        &+& \rhoc\E + \j\times\B\; .
\label{consmom}
\endeqar
Here $\epsilon$, $\S$ may be the full expressions for a hot plasma.
The stress-energy space-tensor of the plasma is
\begeq
        \tarrow{T} \equiv (\rho_{\rm m}\pri +
        p\pri)\gamma^2 \v\otimes\v + p\pri\tarrow{h}
	\approx
	\rho_{\rm m}\pri\gamma^2 \v\otimes\v + p\pri\tarrow{h}
\endeq
in consistence with Def.~(\ref{defplas}).
The fact that, in the cold plasma limit, Eqs.~(\ref{consmom}) and 
(\ref{contplas}) can be combined to yield Eq.~(\ref{eqmot1}), without 
explicit use of inequality~(\ref{schldiff}), implies that the collision terms 
have to implicitly guarantee synchronized bulk acceleration of the components 
for Def.~(\ref{defplas}) to be self-consistent in the first place.
In the cold plasma limit, Eq.~(\ref{consmom}) can be used
alternatively to Eq.~(\ref{eqmot1}). This is particularly appropriate for 
numerical treatments of MHD, which requires the solutions of a conservative 
system of differential \equ s.

\subsection{The law of energy conservation}
It is straightforward to show that the laws of energy conservation for 
the individual plasma components 
add linearly to give 
\begeq
	\1d{\ag}\left(\delt{} - \vbeta\cdot\n\right)\epsilon
	= -\div\S + 2 \vec{g}\cdot\S
	+ \tarrow{H}:\tarrow{T} + \j\cdot \E\; .
\endeq
Like Eq.~(\ref{consmom}) this \equ\ is exact, but it makes sense as part of a 
complete set of \equ s only with the 
single-cold-fluid assumptions discussed above.

\subsection{The generalized Ohm's law for an electron-ion plasma}
Multiplying the \equ\ of motion for the ions by \(Z m_{\rm e}\gamma_{\rm e}^2\)
and the \equ\ of motion for the electrons by \(m_{\rm i}\gamma_{\rm i}^2\) and 
subtracting the latter from the former yields
\begeqar
	\lefteqn{ m_{\rm i}m_{\rm e}\left[Z \gamma_{\rm e}^2 n_{\rm i}
	\ddtau{i}{(\gamma_{\rm i}\v_{\rm i})} - \gamma_{\rm i}^2 n_{\rm e}
	\ddtau{e}{(\gamma_{\rm e}\v_{\rm e})}\right] }\nonu\\
	&& = m_{\rm i}m_{\rm e}\gamma_{\rm i}^2\gamma_{\rm e}^2
	     \left[(Z n_{\rm i} - n_{\rm e})\vec{g} 
	+    \tarrow{H}\cdot(Z n_{\rm i}\v_{\rm i}-n_{\rm e}\v_{\rm e})\right]
	\nonu\\
	&& -{}\frac{Z m_{\rm e}\gamma_{\rm e}^2}{\ag}\n(\ag p_{\rm i}\rmi)
	   +\frac{m_{\rm i}\gamma_{\rm i}^2}{\ag}\n(\ag p_{\rm e}\rme)\nonu\\
	&& + {}(Z m_{\rm e}\gamma_{\rm e}^2\rho_{\rm ci} 
	   - m_{\rm i}\gamma_{\rm i}^2\rho_{\rm ce})\E 
	   + (Z m_{\rm e}\gamma_{\rm e}^2\j_{\rm i}  
           - m_{\rm i}\gamma_{\rm i}^2\j_{\rm e})\times\B \nonu\\
        && + {}Z m_{\rm e}\gamma_{\rm e}^2\vec{k}_{\rm i}
	   - m_{\rm i}\gamma_{\rm i}^2\vec{k}_{\rm e}\; .
\label{preOhm}
\endeqar
This \equ\ can be simplified by dropping terms of the order 
$m_{\rm e}/m_{\rm i}$ as compared to their companion terms: 
With \(O(\gamma_{\rm i}^2)=O(\gamma_{\rm e}^2)\) and the assumptions
\(p_{\rm i}\rmi/p_{\rm e}\rme \ll m_{\rm i} / m_{\rm e}\), 
\(\rho_{\rm ci}/\rho_{\rm ce}\ll m_{\rm i} / m_{\rm e}\), the terms with 
$p_{\rm i}\rmi$, $\rho_{\rm ci}$ and $\vec{k}_{\rm i}$ can be neglected.
The l.h.s. of the \equ\ and the \gm\ term are approximated according to 
inequalities~(\ref{schldiff}) and (\ref{syncHv}), respectively. 
Finally the term \( (\#)\times\B\) can be 
manipulated as elaborated in appendix~(\ref{xxB}) to arrive at
\begeqar 
	\lefteqn{
	\frac{\me(Z n\subi\gammae^2  - n\sube\gammai^2)}
		{e n\sube\gammae\gammai^2}\left(\ddtau{p}{(\gamma\v)} 
		- \tarrow{H}\cdot(\gamma^2\v)\right)		}\nonu\\
	&&\approx \E
        +\frac{Z n\subi\gammai}{n\sube\gammae}\v\times\B
        -\frac{\j\times\B}{e n\sube\gammae}
	+\frac{\n(\ag p\sube\rme)}{e n\sube\gammae\ag} \nonu\\
	&&+{}\frac{\me\gammae}{e n\sube}(Z n\subi - n\sube)\vec{g} 
	 - \1d{e n\sube\gammae}\vec{k}\sube \; ,
\label{allgOhm0}
\endeqar
where $\tau_{\rm p}$ is the proper time in the plasma frame $K\pri$.

In analogy to the classical two-component theory and consistent with the 
special relativistic treatment of Blackman \& Field (1993) one can make 
the ansatz (motivated by the fact that $\vec{k}\sube$ is due to the average 
momentum transfer between electrons and ions)
\begeq
	\vec{k}\sube = n\sube\nu_{\rm c}\me
				(\gammai\v\subi - \gammae\v\sube)\; .
\label{ansatzke}
\endeq
The effective electron collision frequency $\nu_{\rm c}$ (electron-ion, 
electron-electron, or electron-wave collisions as measured in the plasma rest 
frame)
may be anisotropic. Classically anisotropy of $\nu_{\rm c}$ is due to the 
reduced mobility of charged particles perpendicular to \mf s. 
It can be expected that, as gravity, the \gm\ force gives rise to drift 
motions and therefore $\nu_{\rm c}$ might depend on the direction relative to 
the gravitomagnetic field as well. Additionally, the \gm\ force causes
gyration of particles around the \gm\ field \(\vec{H}=\ag^{-1}\rot\vbeta \) 
with a gyrofrequency as measured in the FIDO frame, $\sim |\vec{H}|$. 
If particles gyrate freely between collisions, i.e. 
if \(|\vec{H}|\gamma/\nu_{\rm c}\sim (\o/\ag)/(\nu_{\rm c}/\gamma) >> 1\) 
the collision rate should become anisotropic. A rough estimate of this ratio 
will be given below.
Further discussion of an anisotropic collision rate is beyond the scope 
of this paper.

Re-expressing $\gammai\v\subi$ and $\gammae\v\sube$ in Eq.~(\ref{ansatzke}) 
with the aid of Eqs.~(\ref{jivj}) and (\ref{jevj}) yields 
\begeq
	\vec{k}\sube =\nu_{\rm c}\me
			(\j -\rhoc\pri\gamma\v)/e\; .
\endeq
Inserting this into Eq.~(\ref{allgOhm0}), one arrives at the 
{\it Generalized Ohm's law}
\begeqar
	\frac{\j}{\sigma\gammae}&\approx&
	\E +\frac{Z n\subi\gammai}{n\sube\gammae}\v\times\B
        -\frac{\j\times\B}{e n\sube\gammae}
	+\frac{\n(\ag p\sube\rme)}{e n\sube\gammae\ag }
	\nonu\\
        &+&{}\frac{4\pi\gammae}{\o^2_{\rm pe}}\rhoc\pri\vec{g}
	+\frac{\rhoc\pri\gamma\v}{\sigma\gammae}
	\nonu\\
        &-&\frac{4\pi e(Z n\subi\gammae^2  - n\sube\gammai^2)}
               {\o_{\rm pe}^2 \gammae\gammai^2}\left(\ddtau{p}{(\gamma\v)}
               - \tarrow{H}\cdot(\gamma^2\v)\right) \; ,
\label{allgOhm}
\endeqar
where the conductivity 
\begeq
	\sigma = \frac{e^2 n\sube}{\me \nu_{\rm c}}
		\equiv \frac{\omega^2_{\rm pe}}{4\pi\nu_{\rm c}}
\label{defsigma}
\endeq 
and $\omega_{\rm pe}$ is the electron plasma frequency.
The appearance of the convection current \(\rhoc\pri\gamma\v\) is 
compulsory to fulfil the transformation law of the current 4-vector and 
thereby confirms the correctness of the ansatz made for $\vec{k}\sube$ in 
Eq.~(\ref{ansatzke}). 

The generalized Ohm's law derived here contains no `current acceleration' 
term (see appendix~\ref{fullohm}). This is a consequence of 
inequalities~(\ref{schldiff}), which has to be applied here in order to be 
consistent with Eq.~(\ref{eqmot1}), i.e. to stay in the MHD regime. The 
current acceleration term is, however,
important only on time-scales in the plasma frame of the order of the 
collision time of the 
electrons, e.g. switch-on processes. There is, however, something
like an `acceleration current', which, after multiplication with 
\(\sigma\gammae\) and for $\gammae\approx \gammai$, can be written as 
\begeq
	\j_{\rm acc}\approx -\frac{\rhoc\pri}{\nu_{\rm c}}\ddtau{p}{(\gamma\v)}
	\; .
\endeq
A similar term can be found in Ardavan (1976) for the special-relativistic 
case and in Kippenhahn \& M\"ollenhoff (1975) for the non-relativistic case.
Both the `acceleration current' and the `\gm\ current' 
\begeq
        \j_{\rm gm}\approx \frac{\rhoc\pri}{\nu_{\rm c}}
	\tarrow{H}\cdot(\gamma^2\v)
\endeq
appear only in a charged plasma.

\subsubsection{The quasi-neutral limit}
\label{quasi}
In the limit of quasi-neutral plasma \((Z n\subi \approx  n\sube)\) and
\(\gammae\approx\gammai\approx\gamma\) Eq.~(\ref{allgOhm}) reduces to
\begeqar
        \j &\approx&       
	\sigma\gamma(\E +\v\times\B) -\frac{\sigma}{e n\sube}(\j\times\B) 
	+\frac{\sigma}{e n\sube\ag}\n(\ag p\sube\rme) \; ,
\label{allgOhmqn}
\endeqar
which contains all the terms, familiar from the non-relativistic generalized 
Ohm's law.
The relative importance of the Hall-term can be estimated by writing 
\begeq
	\frac{\sigma}{e n\sube c}\, |\j\times\B| = 
	\frac{\o_{\rm ce}}{(\nu_{\rm c}/\gamma)}\, 
			\left|\j\times\frac{\B}{|\B|}\right|\le
	\frac{\o_{\rm ce}}{(\nu_{\rm c}/\gamma)}\, |\j|\; .
\endeq 
This shows that the Hall-term gives significant modification of the current, 
if the electron gyrofrequency measured in the FIDO frame, 
\(\o_{\rm ce}= e |\B|/\gamma\me c\),
is of the order or higher than the electron collision rate.
According to Spitzer (1962) in a non-magnetized plasma
\begeq
	\nu_{\rm c}\approx 1.5\cdot 10^{-6}\left(\frac{n\sube}{\cm^{-3}}\right)
		\left(\frac{T}{\eV}\right)^{-3/2} \ln\Lambda\, \s^{-1}\; ,
\endeq
where $\ln\Lambda$ is the Coulomb logarithm.
For polytropic ($\Gamma=5/3$) spherical accretion onto a non-rotating 
black hole (the solutions in Kerr metric will be similar) we have 
(e.g. Shapiro \& Teukolski 1983)
\begeq
	n\sube \approx 2\cdot 10^{15} M_1^{-1} \frac{\dot m}{0.01} 
		\left(\frac{r}{\rg}\right)^{-3/2} \cm^{-3}
\label{nespheraccr}
\endeq
\begeq
	T\approx 2\cdot 10^{12}\left(\frac{r}{\rg}\right)^{-1} {\rm K}\; ,
\endeq
with \(\rg = GM/c^2\) and \(M_1\equiv M/10\Msol\).
The dimensionless accretion rate $\dot m$ is measured in units of the 
Eddington rate $\dot M_{\rm Edd} \equiv L_{\rm Edd}/ c^2$. 
Allowing for cooling, the temperature of a realistic accretion flow will 
rather be $\sim 10^{10}{\rm K}$. Thus we have 
\begeq
        \nu_{\rm c} \approx 3\cdot 10^1\, M_1^{-1} \frac{\dot m}{0.01}
		\frac{\ln\Lambda}{10}\, \s^{-1}
\endeq
and
\begeq
	\frac{\o_{\rm ce}}{(\nu_{\rm c}/\gamma)}\approx 6\cdot 10^2\, M_1
		\left(\frac{|\B|}{\rm mGauss}\right)
		\left(\frac{\dot m}{0.01}\right)^{-1}
		\left(\frac{\ln\Lambda}{10}\right)^{-1}\; .
\endeq
Obviously, at those high temperatures and low collision rates, the Hall-term 
would dominate the current even at low field strength. Note, however, that 
in such a situation cross-field driven plasma instabilities are likely to 
occur, resulting in an increased anomalous collision frequency. 

As was mentioned above, also \gm\ drifts might cause anisotropic collision 
rates. The \gm\ gyrofrequency is given by 
\begeq
	|\vec{H}|\sim\o = \frac{c}{2\rg}\left(\frac{2\rg\o}{c}\right)
		= 10^4\, M_1^{-1}\left(\frac{2\rg\o}{c}\right)\, \s^{-1}\; .
\endeq
The ratio of \gm\ gyrofrequency to Spitzer's collision rate is, as measured 
in FIDO frames,
\begeq
	\frac{(\o/\ag)}{(\nu_{\rm c}/\gamma)}\sim 3\cdot 10^2\,
		\frac{\gamma}{\ag}
		\left(\frac{2\rg\o}{c}\right)
		\left(\frac{\dot m}{0.01}\right)^{-1}
                \left(\frac{\ln\Lambda}{10}\right)^{-1}\; .
\label{otonuc}
\endeq
This result hints to a problem: 
$\o$ is the inverse of a dynamical timescale. By applying the
MHD-approximation (Eqs.~[\ref{schldiff}] and [\ref{syncHv}]), which has led to 
neglecting the
current acceleration term, we have implicitly made the assumption that
our timescales of interest are much larger than the electron collision time.
At very high temperatures Spitzer's collision rate is therefore too low
to facilitate single-fluid behaviour of the plasma. 
Thus the \gm\ field should introduce anisotropy of $\nu_{\rm c}$ and of the
particle distribution function. An anisotropic distribution function could 
trigger plasma instabilities with anomalous collision rates, which should 
make \((\o/\ag)/(\nu_{\rm c}/\gamma) < 1\). 
At the horizon the ratio of locally measured \gm\ gyro- to collision frequency 
blows up as \(\gamma/\ag \rightarrow\infty\), which means that the MHD 
description breaks down for FIDOs, i.e. the plasma appears to behave as test 
particles.

\subsection{Summary of \equ s}
As was argued above, a closed set of \equ s of MHD can only be formulated 
in the cold plasma limit with \(\gammae\approx\gammai\approx\gamma\). 
The complete set of \equ s for a electron-ion plasma are:\\
The continuity \equ\
\begeq
        \left(\delt{} - \vbeta\cdot\n\right)
        (\rhom\pri\gamma) + \div(\ag\rhom\pri\gamma\v) = 0\; ,
\label{sumpartcon}
\endeq
with \(\rhom\pri\approx n\subi\mi\). \\
The local law of momentum conservation
\begeqar
        \1d{\ag}\left(\delt{} - \vbeta\cdot\n\right)\vec{S}&=&
     \epsilon\vec{g} + \tarrow{H}\cdot\vec{S}
        - \1d{\ag}\div(\ag\tarrow{T})   \nonu\\
        &+& \rhoc\E + \j\times\B\; ,
\label{sumconsmom}
\endeqar
with $\epsilon\approx\rhom\pri\gamma^2$, $\S \approx \rhom\pri\gamma^2\v$ and 
\(\tarrow{T} \approx \rho_{\rm m}\pri\gamma^2 \v\otimes\v + p\pri\tarrow{h}\).
The generalized Ohm's law 
\begeqar
        \frac{\j}{\sigma\gamma}&\approx&
        \E +\frac{Z n\subi}{n\sube}\v\times\B
        -\frac{\j\times\B}{e n\sube\gamma}
        +\frac{\n(\ag p\sube\rme)}{e n\sube\gamma\ag }
        \nonu\\
        &+&{}\frac{4\pi\gamma}{\o^2_{\rm pe}}\rhoc\pri\vec{g}
        +\frac{\rhoc\pri\gamma\v}{\sigma\gamma}
        \nonu\\
        &-&\frac{4\pi \rhoc\pri}
               {\o_{\rm pe}^2 \gamma}\left(\ddtau{p}{(\gamma\v)}
               - \tarrow{H}\cdot(\gamma^2\v)\right)
                 \; .
\label{sumallgOhm}
\endeqar
Note that \(n\sube \approx Z \rhom\pri/\mi - \rhoc\pri/e\) and 
$\rhoc\pri$ has to be calculated from Eq.~(\ref{transrho}).
The electron pressure follows, e.g., with the assumption that electrons and 
ions have the same temperature. Then \(p\sube\rme\approx Z p\pri / (Z+1)\).\\
The local law of energy conservation
\begeq
        \1d{\ag}\left(\delt{} - \vbeta\cdot\n\right)\epsilon
        = -\div\S + 2\vec{g} \cdot\S
        + \tarrow{H}:\tarrow{T} + \j\cdot\E\; .
\label{sumencons}
\endeq
The law of charge conservation
\begeq
        \left(\delt{} - \vbeta\cdot\n\right)
        \rho_{\rm c} + \div(\ag\j) = 0\; .
\label{sumcharcon}
\endeq
The set of \equ s is completed by Maxwell's \equ s (Thorne et al. 1986), 
from which $\rho_{\rm c}$ and $\j$ can be computed without having to solve 
for $\v\subx$ in the two-fluid theory:
\begeq
	\div \B = 0\; , \qquad \div\E = 4\pi\rhoc\; , 
\endeq
\begeq
	\rot (\ag\E ) = -\left[ \delt{} -
                                {\cal L}_{\vbeta}\right]\B\; , 
\endeq
\begeq
	\rot (\ag\B ) = \left[ \delt{} -
                                {\cal L}_{\vec{\beta}}\right]\E
                                        + 4\pi\ag\j             \; ,
\endeq
where ${\cal L}_{\vbeta}$ is the Lie\--derivative along $\vbeta\; ,$ e.g.
\({\cal L}_{\vbeta}\E = (\vbeta\cdot\n)\E - (\E\cdot\n){\vbeta}\).

\subsubsection{Applicability of MHD and quasi-neutrality}
\label{subsumm}
I have mentioned criteria that limit the applicability of MHD: 
(i) Relativistically hot plasma and 
(ii) if the collision timescale of electrons becomes longer than dynamical 
timescales. The latter happens for low densities and high temperature, unless 
the collision rate is anomalous, or in a FIDO frame close to the horizon where 
$\gamma/\nu_{\rm c}\rightarrow\infty$. As a muon gas that becomes 
`decayless', if it moves with high Lorentz factor through an observer's frame, 
a plasma becomes collisionless in the frame of a FIDO close to the horizon. 
If case (ii) applies (e.g. if the Debye length becomes very large) the 
assumption of quasi-neutrality should also break down.

\section{Conclusion}
The full set of MHD equations, including a generalized Ohm's law, for an 
inviscid fully ionized electron-ion plasma have been derived from a 
two-component plasma theory within the framework of the 3+1 split of the Kerr 
metric. It has been argued that the plasma can only be described as a single 
fluid, if its components are cold (i.e. specific internal energy density 
and pressure are
negligible compared to the rest-mass energy density). This places constraints 
on the applicability of not only MHD but also of hydrodynamics to astrophysical 
objects such as hot relativistic jets and the hot regime of transsonic 
accretion flows onto black holes. Very close to the horizon FIDOs can no 
longer describe a plasma as a fluid, because the plasma particles 
become collisionless on local dynamical timescales.

The generalized Ohm's law is found to be equivalent to the 
special-relativistic expression, if the plasma is quasi-neutral in its rest 
frame. Gravito\magn\ effects of the order of `current acceleration' 
terms appear explicitly only in charged plasma, or if the MHD-approximation 
of synchronized bulk and gravitomagnetic acceleration of the plasma 
components is not applied.

Unlike Ardavan (1976) and Blackman \& Field (1993), who have derived their
special-relativistic generalized Ohm's laws from invariant distribution 
functions, I have derived the generalized Ohm's law by combining the \equ s 
of motion for electrons and ions, thereby compromising 
details of the interaction between electrons and ions. 
The ansatz made for the collision term is, however, confirmed by yielding, 
relativistically correctly, the conduction current.

In a companion paper (Khanna 1998) I will apply this theory to describe a
\gm\ battery, which operates in plasma surrounding a Kerr black hole.

\section*{Acknowledgements}
I gratefully acknowledge fruitful discussions with Max Camenzind, Harald Lesch, 
Jochen Peitz, Stefan Spindeldreher and Markus Thiele. This work is supported 
by the Deutsche Forschungsgemeinschaft (SFB 328).

\appendix
\section{Transformation of charge densities}
\label{apptransrho}
Charge densities in the FIDO frame are given by 
\(\rho_{\rm cx} = -J^{\mu}_{\rm x}U_{\mu}\). 
The transformation of charge densities between FIDO frame and some other frame 
follows from expressing the FIDO 4-velocity $U_{\mu}$ in terms of the 
other frame's 4-velocity $W_{{\rm o}\mu}$ and its velocity as 
measured by FIDOs $v_{{\rm o}\mu}$ (see Khanna \& Camenzind 1996a). The 
transformation between FIDO frame and plasma rest frame $K\pri$ is given by 
\begeq
	\rho_{\rm cx} = -J^{\mu}_{\rm x}
	\left(\frac{W_{\mu}}{\gamma} - v_{\mu}\right)
	=\frac{\rho_{\rm cx}\pri}{\gamma} 
	+ J^{\mu}_{\rm x}v_{\mu} 
	= \frac{\rho_{\rm cx}\pri}{\gamma} + \j_{\rm x}\cdot\v
	\; .
\label{transrho}
\endeq
The last equality follows by applying the definition of the current density 
measured by FIDOs \(j^{\mu}_{\rm x} = h^{\mu}_{\ \nu}J^{\nu}_{\rm x}\):
\begeq
	J^{\mu}_{\rm x}v_{\mu} = \rho_{\rm cx} U^{\mu} v_{\mu} 
	+ j^{\mu}_{\rm x} v_{\mu} 
	= j^{\mu}_{\rm x} v_{\mu}\; .
\endeq
Here the subscript `x' may be `i' for ions, `e' for electrons or {\it blank} 
for the plasma. Inserting Eqs.~(\ref{jie}) into Eq.~(\ref{transrho}) yields 
the second equalities in Eqs.~(\ref{rhocie}). The transformations between 
FIDO frame and electron rest frame $K\rme$, or ion rest frame $K\rmi$, 
respectively, follow by inserting 
\(U_{\mu}=W_{{\rm x}\mu}/\gamma_{\rm x} - v_{{\rm x}\mu}\) into 
\(\rho_{\rm cx} = -J^{\mu}_{\rm x}U_{\mu}\), where now x=i,e. This yields 
\(\rho_{\rm cx} = \rho_{\rm cx}\rmx \gamma_{\rm x}\).

\section{Particle conservation in the FIDO frame}
\label{appconspart}
Particle conservation is expressed in the spacetime viewpoint by
\(N_{{\rm x}\, ;\a}^{\a}=0\), where the particle current 
\(N\subx^{\a}= n_{\rm x} W^{\a}_{\rm x}\) and $ W^{\a}_{\rm x}$ is the 
4-velocity of the species `x'. \(N\subx^{\a}\) splits into 
\begeq
	n\subx^{\a}=h^{\a}_{\ \nu}N\subx^{\nu}
	=n_{\rm x}\gamma_{\rm x} v^{\nu}_{\rm x}
\hbox{   and   }
        \rho_{\rm nx}=-U_{\nu}N\subx^{\nu}=n\subx\gamma_{\rm x} \; .
\endeq
Therefore
\begeq
        N_{{\rm x}\, ;\a}^{\a} = (\rho_{\rm nx}U^{\a}+n\subx^{\a})_{;\a} 
	= 0
	\; ,
\endeq
which leads, under observance of the relation for the determinants of 
4-metric and 3-metric \(g = -\ag^2 h \) and multiplication with $\mx$, 
to Eq.~(\ref{continuity}).

\section{Synchronized bulk acceleration of ions and electrons}
\label{appschldiff}
Adding Eqs.~(\ref{eqmotx}) (x=i,e) and using definition (\ref{bulkacc}) yields 
on the l.h.s.
\begeqar
\lefteqn{\rho_{\rm mi}\rmi\ddtau{i}{(\gamma_{\rm i}\v_{\rm i})} 
	+\rho_{\rm me}\rme\ddtau{e}{(\gamma_{\rm e}\v_{\rm e})}
	=\rho_{\rm mi}\rmi\ddtau{i}{(\gamma_{\rm i}\v_{\rm i})} }\nonu\\
	&&+ \rho_{\rm me}\rme\left(\ddtau{i}{(\gamma_{\rm i}\v_{\rm i})} 
		- \left(\ddtau{i}{(\gamma_{\rm i}\v_{\rm i})}
		-\ddtau{e}{(\gamma_{\rm e}\v_{\rm e})}\right)\right)\nonu\\
	&&\approx(\rhomi\rmi+\rhome\rme)
				\ddtau{i}{(\gamma_{\rm i}\v_{\rm i})}
	  \approx\rho_{\rm m}\pri\ddtau{p}{(\gamma\v)} \; ,
\endeqar
which is the l.h.s. of Eq.~(\ref{eqmot1}).
The following assumptions have been made in order to arrive at the last 
line: (i) Inequation~(\ref{schldiff}) has been used and
(ii) approximation~(\ref{gamma2v}) has been applied.

\section{Manipulating \((Z \me\gammae^2\j\subi - \mi\gammai^2\j\sube)\)}
\label{xxB}
Equation~(\ref{preOhm}) contains the term
\((Z \me\gammae^2\j\subi - \mi\gammai^2\j\sube)\times\B\). In 
Eq.~(\ref{allgOhm}) this term has mutated into 
\(Z e n\subi\mi\gamma^2(\gammai\v - \j / Z e n\subi)\times\B\). 
This happened in the following way: The \equ s
\begeq
	\j = \j\subi + \j\sube = \rhoci\v\subi + \rhoce\v\sube
\endeq 
and 
\begeq
        \gamma^2\rhom\pri\v \approx \gammai^2\rhomi\rmi\v\subi + 
			      \gammae^2\rhome\rme\v\sube
\endeq 
can be combined to give
\begeq
	\mi\gammai^2\j\sube\approx\mi\gammai^2\j 
			- Z e \gammai\gamma^2\rhom\pri\v
\label{jevj}
\endeq
and
\begeq
        Z\me\gammae^2\j\subi\approx Z\rhoci\me\gammae^2\v 
	-\frac{Z\me\rhome\rme\gammae^4\rhoci}{\rhomi\rmi\gammai^2\rhoce}\j\; ,
\label{jivj}
\endeq
where approximations of the order $\me / \mi$ and as given in 
Eq.~(\ref{gamma2v}) have been made. Putting things together and making the 
same kind of approximations yields the desired result.

\section{Generalized Ohm's law without synchronized acceleration}
\label{fullohm}
If the approximations of synchronized bulk and \gm\ acceleration 
(Eqs.~[\ref{schldiff}] and [\ref{syncHv}]) are not applied to 
Eq.~(\ref{preOhm}), and if $\gammax\v\subx$ in the bulk acceleration terms  
and $\v\subx$ in the \gm\ force term are eliminated with the aid of 
Eqs.~(\ref{jevj}) and (\ref{jivj}), one gets instead of the l.h.s. of 
Eq.~(\ref{allgOhm0}) (or instead of the last line of Eq.~[\ref{allgOhm}] 
with opposite sign) 
\begeqar
\lefteqn{
	\frac{Z n\subi\gammae\me}{n\sube e \gammai^2}\left[
	\frac{\gammai}{\ag}\delt{} + \left(\gamma\v -\frac{\gammai}{\ag}
	\vbeta\right)\cdot\n\right](\gamma\v)}	\nonu\\
	&&-\frac{\me}{e\gammae}\left[\frac{\gammae}{\ag}\delt{} + 
	\left(\frac{Z n\subi}{n\sube}\gamma\v -\frac{\j}{e n\sube}
	-\frac{\gammae}{\ag}\vbeta\right)\cdot\n\right]
	\frac{Z n\subi}{n\sube}\gamma\v\nonu\\
	&&+\frac{\me}{e\gammae}\left[\frac{\gammae}{\ag}\delt{} + 
	\left(\frac{Z n\subi}{n\sube}\gamma\v -\frac{\j}{e n\sube}
	-\frac{\gammae}{\ag}\vbeta\right)\cdot\n\right]
	\frac{\j}{e n\sube}\nonu\\
	&&-\frac{\me}{e\gammae}\frac{\gammae}{\ag}\n\vbeta\cdot
	\left(\frac{Z n\subi}{n\sube}(\gammae-\gammai)\v
	+\frac{\j}{e n\sube}\right)\; .
\label{lhsOhm}
\endeqar
One can recognize acceleration terms, current acceleration terms and 
gravitomagnetic terms as well.
If the plasma is quasi-neutral and for \(\gammae\approx\gammai\approx\gamma\), 
the first line and the second line (except \(\j\cdot\n(\gamma\v)\)) cancel.

It is important to note that, any extra term containing $\j$ or $\B$ in 
Ohm's law may allow for dynamo action. In particular, if such a term 
possessed an axisymmetric toroidal component that does not vanish in a neutral 
point of an axisymmetric \po\ \mf\ loop, it might support the growth of 
axisymmetric magnetic field.
This would violate Cowling's theorem (Cowling 1934; Khanna \& Camenzind 1996a).

The \gm\ terms \((\vbeta\cdot\n)\j/e n\sube\) and \(\n\vbeta\cdot\j/e n\sube\) 
do, in fact, individually possess such terms 
\((\pm \o (\j_{\rm p}\cdot\n\ln\ot)\ephi/e n\sube)\), but they cancel 
identically. 
If inequalities~(\ref{schldiff}) 
(approximation of synchronized bulk acceleration) 
were applied to Eq.~(\ref{preOhm}), without 
approximating the \gm\ force terms in the same way (Eq.~[\ref{syncHv}]), the 
only \gm\ term in $\j$ that remained, would be the $\n\vbeta\cdot\j$-term of 
Eq.~(\ref{lhsOhm}). This would allow for 
an axisymmetric \gm\ \dy , but would be an artefact of an incomplete 
approximation.
The generalized Ohm's law does therefore not introduce a new 
axisymmetric {\it \gm} \dy . 

The current acceleration terms are of the order 
\begeq
	\frac{4\pi}{\o_{\rm pe}^2\gammae}\frac{d\j}{d\tau_{\rm p}}\; .
\endeq 
They will, in general, possess axisymmetric toroidal components, but 
compared to the conduction current $\j/\sigma\gammae$ they are of the order 
\((\nu_{\rm c}\tau_{\rm p})^{-1}\), i.e. they violate Cowling's theorem only on 
timescales of the electron collision timescale, where the applicability of MHD 
breaks down. 


\begin{thebibliography}{}
\bibitem [\protect\citename{Ardavan }1976]{Ard} Ardavan, H., 1976, ApJ 203, 226
\bibitem [\protect\citename{Bekenstein \& Eichler }1985]{BekEi} Bekenstein, J., 
Eichler, D., 1985, ApJ 298, 493
\bibitem [\protect\citename{Bekenstein \& Oron }1978]{BekOr} Bekenstein, J., 
Oron, E., 1978, Phys. Rev. D 18, No. 6, 1809
\bibitem [\protect\citename{Beskin \& Par'ev }1993]{BesPa} Beskin, V.S., 
Par'ev, V.I., 1993, Phys. Uspekhi 36 (6), 529
\bibitem [\protect\citename{Blackman \& Field }1993]{BlaFie} Blackman, E.G., 
Field, G.B., 1993, Phys. Rev. Let. 71, No. 21, 3481
\bibitem [\protect\citename{Blandford \& Znajek }1977]{BlaZna} Blandford, R.D., 
Znajek, R.L., 1977, MNRAS 179, 433 
\bibitem [\protect\citename{Camenzind } 1986]{Ca96} Camenzind, M., 1986, A\&A 
162, 32
\bibitem [\protect\citename{Camenzind } 1987]{Ca97} Camenzind, M., 1987, A\&A 
184, 341
\bibitem[\protect\citename{Cowling }1934]{Cow34} Cowling T.G., 1934, MNRAS 94,
39
\bibitem [\protect\citename{Fendt }1997]{Fe} Fendt, C., 1997, A\&A, 319, 1025
\bibitem [\protect\citename{Khanna }1997]{Kha97} Khanna, R., 
1997, in: Proceedings of the Second International Sakharov Conf. on 
Phys., Dremin, I.M., Semikhatov, A.M. (Eds.), World Scientific, p. 134
\bibitem [\protect\citename{Khanna }1998]{Kha98} Khanna, R., 1998, 
MNRAS, in press 
\bibitem [\protect\citename{Khanna \& Camenzind }1994]{KhaCa94} Khanna, R., 
Camenzind, M., 1994, ApJ 435, L129
\bibitem [\protect\citename{Khanna \& Camenzind }1996a]{KhaCa96a} Khanna, R., 
Camenzind, M., 1996a, A\&A 307, 665
\bibitem [\protect\citename{Khanna \& Camenzind }1996b]{KhaCa96b} Khanna, R., 
Camenzind, M., 1996b, A\&A 313, 1028
\bibitem [\protect\citename{Kippenhahn \& M\"ollenhoff }1975]{KiMo} Kippenhahn, 
R., M\"ollenhoff, C., 1975, Elementare Plasmaphysik, 
B.I.-Wissenschaftsverlag
\bibitem [\protect\citename{Kudoh \& Kaburaki }1996]{KuKa} Kudoh, T., 
Kaburaki, O., 1996, ApJ 460, 199
\bibitem [\protect\citename{Okamoto }1992]{Oka} Okamoto, I., 1992, MNRAS 254, 
192
\bibitem [\protect\citename{Phinney }1983]{Phin} Phinney, E.S., Ph.D. diss., 
Univ. of Cambridge
\bibitem [\protect\citename{Schl\"uter } 1950]{Schluter} Schl\"uter, A., 1950,
Z. Naturforschung 5a, 72 
\bibitem [\protect\citename{Shapiro \& Teukolsky }1983]{ShapTeu} Shapiro, S.L., 
Teukolsky, S.A., 1983, Black Holes, White Dwarfs and Neutron Stars, 
Wiley-Interscience, New York
\bibitem [\protect\citename{Spitzer }1962]{Spitzer} Spitzer, L., 1962, 
Physics of Fully Ionized Gases, Interscience, New York
\bibitem [\protect\citename{Takahashi et al. } 1990]{Taketal} Takahashi, M., 
Nitta, S., Tatematsu, Y., Tomimatsu, A., 1990, ApJ 363, 206
\bibitem [\protect\citename{Thorne \& Macdonald }1982]{ThoMac} Thorne, K.S., 
Macdonald, D.A., 1982, MNRAS 198, 339
\bibitem [\protect\citename{Thorne et al. }1986]{TPM} Thorne, K.S., Price, 
R.H., Macdonald, D.A., Suen, W.-M., Zhang, X.-H., 1986, in: Black Holes: 
The Membrane Paradigm, Thorne, K.S., Price, R.H., Macdonald, D.A., (Eds.), 
Yale Univ. Press, New Haven

\end{thebibliography}
\end{document}